\documentclass[12pt,epsf]{article}
\pdfoutput=1
\usepackage[pdftex]{graphicx}
\usepackage{amsmath,amssymb}
\usepackage{bm}
\usepackage{ascmac}
\usepackage{braket}
\graphicspath{{./fig/}}
\usepackage{comment}
\usepackage{cite}

\newcommand{\beq}{\begin{equation}}
\newcommand{\eeq}{\end{equation}}
\newcommand{\bea}{\begin{eqnarray}}
\newcommand{\eea}{\end{eqnarray}}

\newcommand{\ba}{\begin{aligned}}
\newcommand{\ea}{\end{aligned}}
\newcommand{\Slash}[1]{{\ooalign{\hfil/\hfil\crcr$#1$}}}

\def\pe2{p_E^2}

\begin{document}
\setlength{\baselineskip}{0.7cm}
\begin{titlepage}
\begin{flushright}
OCU-PHYS 537  \\
NITEP 97
\end{flushright}
\vspace*{10mm}%
\begin{center}{\LARGE\bf
Nonvanishing Finite Scalar Mass \\
\vspace*{5mm}
in Flux Compactification
}
\end{center}
\vspace*{10mm}
\begin{center}
{\Large Takuya Hirose}$^{a}$ and 
{\Large Nobuhito Maru}$^{a,b}$, 
\end{center}
\vspace*{0.2cm}
\begin{center}
${}^{a}${\it
Department of Mathematics and Physics, Osaka City University, \\
Osaka 558-8585, Japan}
\\
${}^{b}${\it Nambu Yoichiro Institute of Theoretical and Experimental Physics (NITEP), \\
Osaka City University,
Osaka 558-8585, Japan}
\end{center}
\vspace*{1cm}

\begin{abstract}
We study possibilities to realize a nonvanishing finite Wilson line (WL) scalar mass in flux compactification.
Generalizing loop integrals in the quantum correction to WL mass at one-loop, 
we derive the conditions for the loop integrals and mode sums in one-loop corrections to WL scalar mass to be finite.
We further guess and classify the four-point and three-point interaction terms satisfying these conditions.
As an illustration, the nonvanishing finite WL scalar mass is explicitly shown in a six dimensional scalar QED 
 by diagrammatic computation and effective potential analysis. 
This is the first example of finite WL scalar mass in flux compactification. 
\end{abstract}
\end{titlepage}

\section{Introduction}
It has been considered that the hierarchy problem is 
 one of the guiding principles to search for the physics beyond the Standard Model (SM) of particle physics.
In the SM, the quantum correction to the mass of Higgs field is sensitive to 
 the square of the ultraviolet cutoff scale of the theory (for example, Planck scale or the scale of grand unified theory).
Since an experimental values of the Higgs mass is 125 GeV, 
 the solution of the hierarchy problem requires an unnatural fine-tuning of parameters or 
 exploring a new physics beyond the SM at order of TeV scale.
Although the latter approaches have been mainly studied so far, 
 no signature of new physics has been found, 
 which is likely to increase the new physics scale, namely Higgs mass. 
Therefore, the solution seems to be desirable such that the Higgs mass is zero at the classical level 
 and is generated by the quantum effects. 

As one of the approaches to the hierarchy problem, 
 higher dimensional theory with magnetic flux compactification has been studied.
Magnetic flux compactification has been originally studied in string theory \cite{BKLS, IU}
Even in the field theories, flux compactification has many attractive properties: 
 attempt to explain the number of the generations of the SM fermion \cite{Witten}, 
 computation of Yukawa coupling \cite{CIM, 0903, MS}.
Recently, it has been shown that the quantum corrections to the masses of zero-mode of the scalar field 
 induced from extra component of higher dimensional gauge field (called Wilson-line (WL) scalar field) are canceled 
 \cite{B1, Lee, B2, HM, 2-loop,HDO}.
The physical reason of the cancellation is that the shift symmetry from translation in extra spaces forbids the mass term of scalar field.
In that situation, the zero-mode of the scalar field can be identified with 
 Nambu-Goldstone (NG) boson of spontaneously broken translational symmetry.
It is not possible for these results to apply to the hierarchy problem as it stands 
 since the scalar field is also massless at quantum level.
If we identify Wilson-line scalar field with Higgs field, 
 we need some mechanism to generate an explicit breaking term of the translational symmetry in compactified space 
 and the scalar field must be a pseudo NG boson such as pion.

In this paper, we study the possibility to realize nonvanishing finite WL scalar mass in flux compactification.
First, we generalize loop integrals in the quantum correction to WL scalar mass at one-loop.
Then, the conditions for the loop integral and mode sum to be finite are derived.
We further guess and classify the four-point and three-point interaction terms 
 generating the finite one-loop quantum correction to WL scalar mass.
Of these interaction terms, 
 we focus on a simplest interaction term 
 and illustrate the finite quantum correction to the WL scalar mass in a six dimensional scalar QED 
 in two ways: diagrammatic computation or effective potential analysis. 
This is the first example of finite WL scalar mass in flux compactification. 

This paper is organized as follows.
We introduce a six-dimensional theory with flux compactification 
 and derive Kaluza-Klein mass spectrum of scalar field, fermion field and SU(2) gauge field in section \ref{FCandKKmass}.
In section \ref{generalQC}, we generalize the loop integrals of quantum correction to the masses of Wilson-line scalar field.
After deriving the conditions for the quantum correction to be finite, 
 the interaction terms providing finite quantum corrections are classified.
In section \ref{intandfinite}, 
 we focus on an interaction term of all interaction terms classified in section \ref{generalQC} 
 and calculate finite quantum correction to WL scalar mass in a six dimensional scalar QED.
In the last section, we summarize our conclusion.
The property of Hurwitz zeta function is summarized in appendix \ref{HZF}.

\section{Flux Compactification and Kaluza-Klein Mass} \label{FCandKKmass}
In this section, we introduce our setup and summarize the Kaluza-Klein mass spectrum of various fields, 
 which are required for calculating the quantum correction to WL scalar mass. 
\subsection{Flux compactification}
Let us first consider a six-dimensional U(1) gauge theory with a constant magnetic flux.
The six-dimensional spacetime is a product of four-dimensional Minkowski spacetime $M^4$ and two-dimensional torus $T^2$.
For later discussion, let us consider the following Lagrangian
	\begin{align}
	\mathcal{L}_{gauge}&=-\frac{1}{4}F_{MN}F^{MN}, \label{gaugelag} \\
	\mathcal{L}_{scalar}&=-D_{M}\overline{\Phi} D^{M}\Phi, \label{scalarlag}\\
	\mathcal{L}_{fermion}&=i\overline{\Psi}\Gamma^M D_M\Psi, \label{fermionlag}
	\end{align}
where the spacetime index is given by $M,N=0,1,\cdots,6,~\mu,\nu=0,1,2,3,~m,n,=5,6$ respectively 
 and we follow the metric convention as $\eta_{MN}=(-1,+1,\cdots+1)$.
The field strength and the covariant derivative of U(1) gauge field $A_M$ are defined 
 by $F_{MN}=\partial_M A_N-\partial_N A_M$, $D_M=\partial_M-igA_M$ with a gauge coupling constant $g$.
$\Phi$ is a bulk scalar field.
$\Gamma^M$ are six-dimensional gamma-matrices.
\footnote{For more detail convention, see \cite{B2,WessBagger}.}
$\Psi$ is a six-dimensional Weyl spinor with two-component Weyl spinors, 
 which satisfies $\Gamma^7\Psi=-\Psi$ in terms of 
 $\Gamma^7\equiv-\Gamma^0\Gamma^1\Gamma^2\Gamma^3\Gamma^5\Gamma^6$.
$\Psi$ has $\psi$ with a charge $-g$ and $\chi$ with a charge $+g$ and is expressed by
	\begin{align}
	\Psi=\left(\begin{array}{l}
	\psi_{L} \\
	\psi_{R}
	\end{array}\right),~~
	\psi_L=\left(\begin{array}{l}
	\psi \\
	0
	\end{array}\right),~~
	\psi_R=\left(\begin{array}{l}
	0 \\
	\bar{\chi}
	\end{array}\right).
	\end{align}

We introduce the magnetic flux in our model.
The magnetic flux is given by the nontrivial background (or vacuum expectation value (VEV)) 
 of the fifth and the sixth component of the gauge field $A_{5,6}$.
This background must satisfy their classical equation of motion $\partial^m\braket{F_{mn}}=0$.
In flux compactification, the background of $A_{5,6}$ is chosen as
	\begin{align}
	\braket{A_5}=-\frac{1}{2}fx_6,~~~\braket{A_6}=\frac{1}{2}fx_5,
	\end{align}
which introduces a constant magnetic flux density $\braket{F_{56}}=f$ with a real number $f$.
Note that this solution breaks an extra-dimensional translational invariance spontaneously.
Integrating over $T^2$, the magnetic flux is quantized as follows
	\begin{align}
	\frac{g}{2\pi}\int_{T^2}dx_5dx_6\braket{F_{56}}=\frac{g}{2\pi}L^2f=N\in\mathbb{Z},
	\end{align}
where $L^2$ is an area of two-dimensional torus.
In the following, we set $L=1$ for simplicity.

It is useful to define $\partial, z$, and $\phi$ as
	\begin{align}
	\partial \equiv \partial_{z}=\partial_{5}-i \partial_{6}, \quad z \equiv \frac{1}{2}\left(x_{5}+i x_{6}\right), 
	\quad \phi=\frac{1}{\sqrt{2}}\left(A_{6}+i A_{5}\right).
	\end{align}
In terms of these complex coordinates and variables, the VEV of $\phi$ is given by $\braket{\phi}=f\bar{z}/\sqrt{2}$.
We expand $\phi$ around the flux background $\phi=\braket{\phi}+\varphi$, where $\varphi$ is a quantum fluctuation.
To distinguish $\varphi$ from an introduced bulk scalar $\Phi$, we call $\varphi$ Wilson line (WL) scalar field.
Defining the covariant derivatives in the complex coordinates is also useful to obtain Kaluza-Klein (KK) masses later, 
 which are defined as
	\begin{align}
	D&=D_5-iD_6=\partial-\sqrt{2}g\phi=\mathcal{D}-\sqrt{2}g\varphi, \\
	\bar{D}&=D_5+iD_6=\bar{\partial}+\sqrt{2}g\bar{\phi}=\bar{\mathcal{D}}+\sqrt{2}g\bar{\varphi},\\
	\mathcal{D}&=\mathcal{D}_5-i\mathcal{D}_6=\partial-\sqrt{2}g\braket{\phi}, \\
	\bar{\mathcal{D}}&=\bar{\mathcal{D}}_5+i\bar{\mathcal{D}}_6=\bar{\partial}+\sqrt{2}g\braket{\bar{\phi}}.
	\end{align}
Note that $\mathcal{D}_m$ means the covariant derivative with VEV.

Finally, we consider an SU(2) Yang-Mills theory with a constant magnetic flux.
The Lagrangian of Yang-Mills theory is given by
	\begin{align}
	\mathcal{L}_{YM}=-\frac{1}{4}F^a_{MN}F^{aMN}, \label{pureYMlag}
	\end{align}
where $a=1,2,3$ are gauge indices.
The field strength and the covariant derivative of SU(2) gauge field $A^a_M$ are defined 
 by $F^a_{MN}=\partial_M A^a_N-\partial_N A^a_M-ig[A_M,A_N]^a$, $D_MA^a_N=\partial_M A^a_N+g\epsilon^{abc}A^b_MA^c_N$.
$\epsilon^{abc}$ is a totally anti-symmetric tensor of SU(2).
In the case of Yang-Mills theory, we introduce a flux background as
	\begin{align}
	\braket{A_{5}^{1}}=-\frac{1}{2} f x_{6}, \quad\braket{A_{6}^{1}}=\frac{1}{2} f x_{5}, \quad\braket{ A_{5}^{2,3}}=\braket{A_{6}^{2,3}}=0.
	\end{align}
For later convenience, we define the covariant derivatives in the complex coordinates as follows
	\begin{align}
	DX^a&=(D_5-iD_6)X^a=\partial X^a-\sqrt{2}g[\phi,X]^a=\mathcal{D}X^a-\sqrt{2}g[\varphi,X]^a, \\
	\bar{D}X^a&=(D_5+iD_6)X^a=\bar{\partial}X^a+\sqrt{2}g[\bar{\phi},X]^a=\bar{\mathcal{D}}X^a+\sqrt{2}g[\bar{\varphi},X]^a,\\
	\mathcal{D}X^a&=(\mathcal{D}_5-i\mathcal{D}_6)X^a=\partial X^a-\sqrt{2}g[\braket{\phi},X]^a, \\
	\bar{\mathcal{D}}X^a&=(\bar{\mathcal{D}}_5+i\bar{\mathcal{D}}_6)X^a=\bar{\partial}X^a+\sqrt{2}g[\braket{\bar{\phi}},X]^a.
	\end{align}

\subsection{Kaluza-Klein mass spectrum}
\label{KKmass}
To compute one-loop correction to WL scalar mass in flux compactification, 
 we need to derive mass eigenvalues for fields propagating in a loop.
In analogy to the quantum mechanics in magnetic field, 
 we regard the covariant derivative $\mathcal{D}$ and $\bar{\mathcal{D}}$ as creation and annihilation operators by
	\begin{align}
	a=\frac{1}{\sqrt{2gf}}i\bar{\mathcal{D}},~~~a^\dag=\frac{1}{\sqrt{2gf}}i\mathcal{D},
	\end{align}
which satisfy the commutation relation $[a,a^\dag]=1$.
Hereafter, we denote $\alpha=2gf$.

The ground state mode function is determined by $a\xi_{0,j}=0,a^\dag\bar{\xi}_{0,j}=0$, 
 where $j=0,\cdots,|N|-1$ accounts for the degeneracy of the ground state.
Creation operator and annihilation operator acts on mode functions as
	\begin{align}
	a^\dag\xi_{n,j}=\sqrt{n+1}\xi_{n+1,j},~~~a\xi_{n,j}=\sqrt{n}\xi_{n-1,j}, \label{creatannihilate}
	\end{align}
and we can construct the higher mode function $\xi_{n,j}$ in the same way as the harmonic oscillator (in detail, see \cite{highermode})
	\begin{align}
	\xi_{n,j}=\frac{1}{\sqrt{n!}}(a^\dag)^n\xi_{0,j},~~~\bar{\xi}_{n,j}=\frac{1}{\sqrt{n!}}(a)^n\bar{\xi}_{0,j},
	\end{align}
where $n=0,1,2\cdots$ is Landau level.
The higher mode function satisfies an orthonormality condition
	\begin{align}
	\int_{T^2}dx^2\bar{\xi}_{n',j'}\xi_{n,j}=\delta_{n,n'}\delta_{j,j'}.
	\end{align}

\subsubsection{Scalar field}

We decompose \eqref{scalarlag} into a four-dimensional part and an extra-dimensional part (see \cite{B1})
	\begin{align}
	\mathcal{L}_{scalar}=-D_\mu\overline{\Phi} D^\mu \Phi-D_m\overline{\Phi} D^m \Phi. \label{scalarlag2}
	\end{align}
Now, we focus on the second term in \eqref{scalarlag2} and extract mass term
	\begin{align}
	\mathcal{L}_{scalar~mass}&=-\mathcal{D}_m\overline{\Phi} \mathcal{D}^m\Phi \nonumber \\
	&=-\overline{\Phi}\alpha\left(a^\dag a+\frac{1}{2}\right)\Phi.
	\end{align}
Then, the KK mass of scalar field is obtained by
	\begin{align}
	m^2_{scalar}=\alpha\left(n+\frac{1}{2}\right), \label{scalarmass}
	\end{align}
where the fact that $a^\dag a$ is a number operator is used.
Note that the KK mass of WL scalar $\varphi^{2,3}$ induced from SU(2) Yang-Mills theory 
 corresponds to \eqref{scalarmass} 
 in Feynman gauge\footnote{If you have a concern about gauge-fixing term and ghost, see \cite{HM}.}.
Although SU(2) Yang-Mills theory involves a ghost field, 
 the KK mass of the ghost field also agrees with \eqref{scalarmass} in Feynman gauge.

\subsubsection{Fermion field}
We decompose \eqref{fermionlag} as in the case of scalar field (see \cite{B2})
	\begin{align}
	\mathcal{L}_{fermion}=i\overline{\Psi}\Gamma^\mu D_\mu\Psi+i\overline{\Psi}\Gamma^5 D_5\Psi+i\overline{\Psi}\Gamma^6 D_6\Psi, \label{fermionlag2}
	\end{align}
and we focus on the second and the third terms in \eqref{fermionlag2}.
Decomposing these terms in terms of two-component Weyl spinors $\psi$ and $\chi$, 
 the mass terms of fermion field are expressed by
	\begin{align}
	\mathcal{L}_{fermion~mass}&=-\chi(\partial-gf\bar{z})\psi-\bar{\chi}(\bar{\partial}-gfz)\bar{\psi}.
	\end{align}
In this case, there are two pairs of annihilation and creation operators
	\begin{align}
	a_{-}&=\frac{i}{\sqrt{\alpha}}(\partial-gf\bar{z}),~~~a^\dag_{-}=\frac{i}{\sqrt{\alpha}}(\bar{\partial}+gf\bar{z}), \\
	a_{+}&=\frac{i}{\sqrt{\alpha}}(\bar{\partial}-gf\bar{z}),~~~a^\dag_{+}=\frac{i}{\sqrt{\alpha}}(\partial+gf\bar{z}),
	\end{align}
where $a_{-},a^\dag_{-}$ act on $\psi$ and $a_{+},a^\dag_{+}$ act on $\chi$.
Using these annihilation and creation operators, we obtain the mass-squared operators for $\psi$ and $\chi$
	\begin{align}
	m^2_{\psi}=\alpha a^\dag_{-}a_{-},~~~m^2_{\chi}=\alpha(a^\dag_{+}a_{+}+1). \label{chiralmass}
	\end{align}
Note that \eqref{chiralmass} means an existence of chiral fermion in flux compactification 
 since zero-mode of $\psi$ is massless but zero-mode of $\chi$ is massive.
We can rewrite \eqref{fermionlag2} in terms of Dirac fermion $\psi_{Lj}$ and $\Psi_{n,j}$,
	\begin{align}
	\psi_{Lj}=\left(\begin{array}{cc}
	\psi_{0,j} \\
	0
	\end{array}\right),~~~
	\Psi_{n,j}=\left(\begin{array}{cc}
	\psi_{n+1,j} \\
	\bar{\chi}_{n,j}
	\end{array}\right),
	\end{align}
and obtain
	\begin{align}
	\mathcal{L}_{fermion~mass}=\sqrt{\alpha(n+1)}\overline{\Psi}_{n,j}\Psi_{n,j}.
	\end{align}
We conclude that $\psi_{Lj}$ is massless and the KK mass of fermion $\Psi_{n,j}$ is given by
	\begin{align}
	m^2_{fermion}=\alpha(n+1). \label{fermionmass}
	\end{align}

\subsubsection{SU(2) gauge field}
Decomposing \eqref{pureYMlag} and focusing on $F^a_{\mu5}F^{a\mu5}+F^a_{\mu6}F^{a\mu6}$ terms (see\cite{HM}), 
 we obtain a mass term for an SU(2) gauge field,
	\begin{align}
	\mathcal{L}_{mass}&=-\frac{1}{2}\partial A_\mu^a\bar{\partial}A^{a\mu}
	+g^2[A_\mu,\braket{\phi}]^a[A^\mu,\braket{\bar{\phi}}]^a
	-\frac{g}{\sqrt{2}}\Big\{-\partial A_\mu^a[A^\mu,\braket{\bar{\phi}}]^a+\bar{\partial}A^a_\mu[A^\mu,\braket{\phi}]^a\Big\} 
	\nonumber \\
	&=-\frac{1}{2}A^a_\mu[-\mathcal{D}\bar{\mathcal{D}}]A^{a\mu}. \label{deform}
	\end{align}
Diagonalizing the covariant derivatives, we find the KK mass of the SU(2) gauge field
	\begin{align}
	m^2_{YM}=\alpha\left(\begin{array}{ccc}
	n_1 & 0 & 0 \\
	0 & n_2 & 0 \\ 
	0 & 0 & n_3+1 \\
	\end{array} \right).\label{YMmass}
	\end{align}
Note that Abelian gauge field is not expressed such as \eqref{deform} 
 since the part of commutator $[A_M, A_N]^a$ is absent in the case of Abelian gauge theory.
Therefore, the KK spectrum of Abelian gauge field is nothing but an ordinary KK mass spectrum 
 $m^2_{U(1)}=(n/R)^2+(m/R)^2$ ($n, m$ is integer and $R=L/(2\pi)$).

\section{Analysis on the divergence structure of loop integral 
and classification of interaction terms} \label{generalQC}
In this section, we systematically analyze the divergence structure of the quantum corrections 
 to WL scalar mass and classify possible interactions providing a finite mass. 

\subsection{The divergence structure of loop integral: part 1}
	\begin{figure}[http]
	\begin{center}
	\includegraphics[width=90mm]{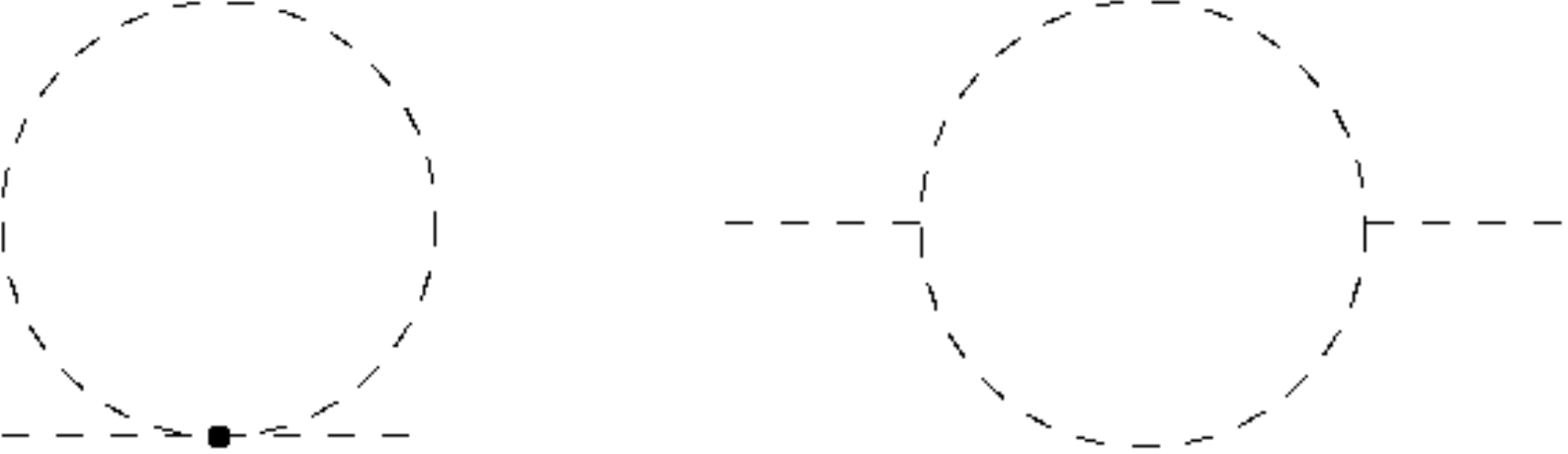}
	\end{center}
	\caption{Example of quantum corrections to the WL scalar mass at one-loop. 
	The fields in the loop are not specified in this figure.}
	\label{fig:one}
	\end{figure}

In this subsection, we investigate the divergence structures for quantum correction to the WL scalar mass at one-loop.
In general, there are two types of Feynman diagrams in figure \ref{fig:one}.
From these diagrams and the results of section \ref{KKmass}, 
 the general form of loop integral in the quantum correction can be written as
	\begin{align}
	I(x;a,b)&=\sum_{n=0}^\infty\int \frac{d^{4} k}{(2 \pi)^{4}} \frac{k^{2a}}{\left(k^{2}+\alpha\left(n+x\right)\right)^b} \nonumber \\
	&=\frac{1}{\alpha^{b-a}}\left(\frac{4\pi}{\alpha}\right)^{\epsilon-2}\frac{\Gamma\left(a+2-\epsilon\right) 
	\Gamma\left(\epsilon+b-a-2\right)}{\Gamma(b) \Gamma\left(2-\epsilon\right)}\zeta[\epsilon+b-a-2,x], 
	\label{Qcorrection}
	\end{align}
where the dimensional regularization was employed for loop integral in the second line.
$\Gamma(z)$ is a gamma function, $\zeta[s,a]$ is Hurwitz zeta function which is summarized in appendix \ref{HZF}, 
 and $d=4-2\epsilon$ dimensions.
$x$ is the part of KK mass characterized by the field running in the loop.
$x=1/2$ corresponds to the KK mass of scalar field \eqref{scalarmass}.
$x=1$ mainly corresponds to the KK mass of fermion field \eqref{fermionmass}.
$x=0$ mainly corresponds to the KK mass of SU(2) gauge field \eqref{YMmass}.
$a$ denotes the number of derivatives acting on the single field and $b$ corresponds to the number of the propagator.
From our interest, we focus on $b=1$ or $b=2$ since we consider  one-loop correction to WL scalar mass.

Since the WL scalar cannot have a bare mass term, 
  the loop integral and mode sum for one-loop correction to WL scalar mass must be finite to realize nonvanishing finite WL scalar mass.
To clarify this point, we investigate
	\begin{align}
	J(x;a,b)\equiv\frac{\Gamma\left(a+2-\epsilon\right) 
	\Gamma\left(\epsilon+b-a-2\right)}{\Gamma\left(2-\epsilon\right)}\zeta[\epsilon+b-a-2,x] \label{Jxab}
	\end{align}
in \eqref{Qcorrection}.
For one-loop corrections, it is enough to consider the case $b=1$ or $b=2$.
In the case of $b=1$, the Gamma function part of \eqref{Jxab} is expressed by
	\begin{align}
	\frac{\Gamma\left(a+2-\epsilon\right) \Gamma\left(\epsilon-a-1\right)}{\Gamma\left(2-\epsilon\right)}
	&=(a+2-\epsilon-1)(a+2-\epsilon-2)\cdots(2-\epsilon)\Gamma(\epsilon-a-1) \nonumber \\
	&=(-1)^a\Gamma(\epsilon-1). \label{gammacal1}
	\end{align}
Thus, $J(x;a,1)$ becomes
	\begin{align}
	J(x;a,1)=(-1)^a\Gamma(\epsilon-1)\zeta[\epsilon-a-1,x].\label{Jxa1}
	\end{align}
In the case of $b=2$, the same part of \eqref{Jxab} is expressed by
	\begin{align}
	\frac{\Gamma\left(a+2-\epsilon\right) \Gamma\left(\epsilon-a\right)}{\Gamma\left(2-\epsilon\right)}
	&=(a+2-\epsilon-1)(a+2-\epsilon-2)\cdots(2-\epsilon)\Gamma(\epsilon-a) \nonumber \\
	&=(-1)^{a}(\epsilon-a-1)\Gamma(\epsilon-1). \label{gammacal2}
	\end{align}
Thus, $J(x;a,2)$ becomes
	\begin{align}
	J(x;a,2)=(-1)^a(\epsilon-a-1)\Gamma(\epsilon-1)\zeta[\epsilon-a,x]. \label{Jxa2}
	\end{align}
Here, Gamma function and Hurwitz zeta function can be expanded in $\epsilon$
	\begin{align}
	\Gamma(\epsilon-1)&=\frac{\Gamma(\epsilon)}{\epsilon-1}=-\left(\frac{1}{\epsilon}-\gamma_E+1+\mathcal{O}(\epsilon)\right), 
	\label{gammaexpand} \\
	\zeta[\epsilon-p,x]&=\zeta[-p,x]+\zeta^{(1,0)}[-p,x]\epsilon+\mathcal{O}(\epsilon^2), \label{zetaexpand}
	\end{align}
where $\gamma_E=0.5772\cdots$ is the Euler-Mascheroni constant, 
 $p$ is an arbitrary positive integer, and $\zeta^{(1,0)}[s,a]=\partial\zeta[s,a]/\partial s$.
Using \eqref{gammaexpand}, \eqref{zetaexpand} and \eqref{finitekey} in appendix \ref{HZF}, 
 the condition for $\Gamma(\epsilon-1)\zeta[\epsilon-p,x]$ being finite is that $p$ is even,
	\begin{align}
	\Gamma(\epsilon-1)\zeta[\epsilon-p,x]=\mathrm{finite},~~~\mathrm{if}~p=even. \label{finiteimage}
	\end{align}
Applying this result to \eqref{Jxa1} and \eqref{Jxa2}, $J(x;a,1)$ takes finite value at odd $a$, 
 $J(x;a,2)$ does at even $a$.

\subsection{Classification of interaction terms: part 1} \label{IQC}
We classify the interaction terms providing finite correction to WL scalar mass at one-loop.
We consider interaction terms which has no derivatives acting on $\varphi$ or $\bar{\varphi}$ 
 because we consider one-loop corrections to WL scalar mass.

\subsubsection{Four-point interaction}

Four-point interaction term generates a correction to WL scalar mass of the left one in figure \ref{fig:one}.
The diagram corresponds to $J(x;a,1)$ ($a$: odd), from which we can guess the four-point interaction terms as follows,
	\begin{itemize}
	\item scalar field loop
		\begin{align}
		J(1/2;a,1)\rightarrow\bar{\varphi}\varphi\partial_{\mu_1}\cdots\partial_{\mu_a}\overline{\Phi}\partial^{\mu_1}\cdots\partial^{\mu_a}\Phi,
		\end{align}
	\item fermion field loop
		\begin{align}
		J(1;a,1)\rightarrow \bar{\varphi}\varphi\bar{\psi}(\Slash{\partial})^{2a-1}\psi, \label{4pointfermi}
		\end{align}
	\item SU(2) gauge field loop
		\begin{align}
		J(0;a,1)\rightarrow\bar{\varphi}\varphi\partial_{\mu_1}\cdots\partial_{\mu_a}A^a_\nu\partial^{\mu_1}\cdots\partial^{\mu_a}A^{a\nu}.
		\end{align}
	\end{itemize}
We did not consider a four-point interaction with such as $\bar{\varphi}\varphi\partial_{\mu_1}\cdots\partial_{\mu_a}\bar{\psi}\partial^{\mu_1}\cdots\partial^{\mu_a}\psi$ since the fermion mass $m_{fermion}=\sqrt{\alpha(n+1)}$ is emerged 
 from a numerator in the fermion propagator and then the form of Hurwitz zeta function is complicated.
On the other hand, $(\Slash{k})^{2a-1}$ is obtained by \eqref{4pointfermi}.
Computing quantum correction, the trace of $\Slash{k}$ from a numerator in the propagator of fermion multiplied 
 by $(\Slash{k})^{2a-1}$ is given by $k^{2a}$.
If $a$ is odd, this term contributes to quantum correction to WL scalar mass.

\subsubsection{Three-point inteaction}
Three-point interaction term generates a correction of the right one in figure \ref{fig:one}.
The diagram corresponds to $J(x;a,2)$ ($a$: even), from which we can guess the three-point interaction terms as follows,
	\begin{itemize}
	\item scalar field loop
		\begin{align}
		J(1/2;0,2)&\rightarrow \bar{\varphi}\overline{\Phi}\Phi+\varphi\overline{\Phi}\Phi, \label{new3pt}\\
		J(1/2;a,2)&\rightarrow \bar{\varphi}\partial_{\mu_1}\cdots\partial_{\mu_{a/2}}\overline{\Phi}
		\partial^{\mu_1}\cdots\partial^{\mu_{a/2}}\Phi+\varphi\partial_{\mu_1}\cdots\partial_{\mu_{a/2}}
		\overline{\Phi}\partial^{\mu_1}\cdots\partial^{\mu_{a/2}}\Phi,
		\end{align}
	\item fermion field loop
		\begin{align}
		J(1;a,2)
		&\rightarrow \bar{\varphi}\bar{\psi}(\Slash{\partial})^{a-1}\psi+\varphi\bar{\psi}(\Slash{\partial})^{a-1}\psi, \label{newfermi}
		\end{align}
	\item SU(2) gauge field loop
		\begin{align}
		J(0;a,2)\rightarrow \bar{\varphi}\partial_{\mu_1}\cdots\partial_{\mu_{a/2}}A^a_\nu 
		\partial^{\mu_1}\cdots\partial^{\mu_{a/2}}A^{a\nu}+\varphi\partial_{\mu_1}\cdots
		\partial_{\mu_{a/2}}A^a_\nu\partial^{\mu_1}\cdots\partial^{\mu_{a/2}}A^{a\nu},
		\end{align}
	\end{itemize}
where $J(1/2;0,2)$ is allowed because of $\zeta[0,1/2]=0$.
We are particularly interested in the interaction term in \eqref{new3pt}, therefore we will discuss in section \ref{intandfinite}.

\subsection{The divergence structure of loop integral: part 2}
In more general, we can consider the interaction term with coefficient depending on KK mode.
The more general form of loop integral in the quantum correction is given by
	\begin{align}
	I'(x;a,b)&=\sum_{n=0}^\infty\int \frac{d^{4} k}{(2 \pi)^{4}} \frac{k^{2a}f(n)}{\left(k^{2}+\alpha\left(n+x\right)\right)^b} \nonumber \\
	&=\frac{1}{(4 \pi)^{d / 2}}\frac{\Gamma\left(a+\frac{d}{2}\right) \Gamma\left(b-a-\frac{d}{2}\right)}{\Gamma(b) 
	\Gamma\left(\frac{d}{2}\right)}\sum_{n=0}^\infty \frac{f(n)}{(\alpha(n+x))^{b-a-\frac{d}{2}}},
	\end{align}
where $f(n)$ is a coefficient generated by an interaction term depending on KK mode $n$.
Since the more complicated the form of $f(n)$ is, the more difficult we express as Hurwitz zeta function, 
 the discussion on the finiteness of loop integral becomes hard to proceed.
As a candidate, $f(n)=((\alpha(n+q))^c$ ($q$ and $c$ are real numbers) is considered.
In this section, we assume $f(n)=\alpha(n+q)$ for simplicity.
 Thus, $I'(x;a,b)$ is expressed by
	\begin{align}
	I'(x;a,b)&=\frac{1}{\alpha^{b-a-1}}\left(\frac{4\pi}{\alpha}\right)^{\epsilon-2}\frac{\Gamma\left(a+2-\epsilon\right) 
	\Gamma\left(\epsilon+b-a-2\right)}{\Gamma(b) \Gamma\left(2-\epsilon\right)} \nonumber \\
	&\quad\quad\quad\quad\times\Big(\zeta[\epsilon+b-a-3,x]+(q-x)\zeta[\epsilon+b-a-2,x]\Big). \label{Qcorrection2}
	\end{align}
If $q\ne x$, the divergence will inevitably appears from either $\zeta[\epsilon+b-a-3,x]$ or $\zeta[\epsilon+b-a-2,x]$. 
To avoid the divergence and see whether the quantum correction is finite, 
 we need to choose $q=x$ (equivalent to the choice $f(n)=$ KK mass), and then investigate
	\begin{align}
	K(x;a,b)\equiv\frac{\Gamma\left(a+2-\epsilon\right) 
	\Gamma\left(\epsilon+b-a-2\right)}{\Gamma\left(2-\epsilon\right)}\zeta[\epsilon+b-a-3,x], \label{Kxab}
	\end{align}
in \eqref{Qcorrection2}.
Substituting $b=1$ or $b=2$ in \eqref{Kxab} and using \eqref{gammacal1} or \eqref{gammacal2}, we obtain
	\begin{align}
	K(x;a,1)&=(-1)^a\Gamma(\epsilon-1)\zeta[\epsilon-a-2,x], \label{Kxa1}\\
	K(x;a,2)&=(-1)^a(\epsilon-a-1)\Gamma(\epsilon-1)\zeta[\epsilon-a-1,x]. \label{Kxa2}
	\end{align}
Applying the result \eqref{finiteimage} to \eqref{Kxa1} and \eqref{Kxa2}, $K(x;a,1)$ takes finite value at even $a$, 
 $K(x;a,2)$ does at odd $a$.

\subsection{Classification of interaction terms: part 2}
We consider the case of four-point interaction terms ($a$: even) and guess their form providing finite quantum corrections to WL scalar mass,
	\begin{itemize}
	\item scalar field loop
		\begin{align}
		K(1/2;0,1)&\rightarrow \bar{\varphi}\varphi\overline{\Phi}\left(a^\dag a+\frac{1}{2}\right)\Phi, \label{new4pt} \\
		K(1/2;a,1)&\rightarrow \bar{\varphi}\varphi\partial_{\mu_1}\cdots\partial_{\mu_a}\overline{\Phi}\left(a^\dag a+\frac{1}{2}\right)\partial^{\mu_1}\cdots\partial^{\mu_a}\Phi,
		\end{align}
	\item fermion field loop
		\begin{align}
		K(1;a,1)\rightarrow \bar{\varphi}\varphi\bar{\psi}(\Slash{\partial})^{2a-1}(a^\dag a+1)\psi
		\end{align}
	\item SU(2) gauge field loop
		\begin{align}
		K(0;a,1)\rightarrow \bar{\varphi}\varphi\partial_{\mu_1}\cdots\partial_{\mu_a}A^a_\nu(a^\dag a)\partial^{\mu_1}\cdots\partial^{\mu_a}A^{a\nu}.
		\end{align}
	\end{itemize}
The case of three-point interaction term is hard to guess 
 because the three-point interaction term cannot be expressed in terms of a mass-squared operator.
Thus, we do not consider the three-point interaction terms in this section.
If we compute loop integral by using $f(n)=(\alpha(n+x))^c$, the meaning of $c$ is the number of mass-squared operators.

One might think that it would be possible that finite quantum corrections from \eqref{new3pt} and \eqref{new4pt} 
 are cancelled if both interactions \eqref{new3pt} and \eqref{new4pt} are present in a theory.
However, we need not worry such a cancellation 
 since the structure of $J(1/2;0,2)=\Gamma(\epsilon)\zeta[\epsilon,1/2]$ is different 
 from $K(1/2;0,1)=\Gamma(\epsilon-1)\zeta[\epsilon-2,1/2]$.


\subsection{Comments on interactions between the field with different KK mode indices}\label{differentKK}
Due to the presence of annihilation and creation operators, 
 there are interactions between the field with different KK mode indices.
For example, they correspond to $\varphi\chi_{n,j}\psi_{n+1,j}$ (see \cite{B2}) 
 and $\varphi A^a_{\mu,n,j}A^{a\mu}_{n+1,j}$ (see \cite{HM}).
The generalization of loop integral with these interaction terms is very complicated 
 because we must use two propagators in different KK modes 
 and cannot systematically identify the coefficient depending on KK mode $n$.

\section{Illustration of nonvanishing finite WL scalar mass} \label{intandfinite}
In section \ref{IQC}, we have classified the interaction terms generating finite quantum correction at one-loop.
Now we focus on \eqref{new3pt} since it has no derivatives and 
 is the simplest interaction term of all interaction terms in section \ref{IQC}.
Therefore, we explicitly calculate finite quantum corrections to WL scalar mass from \eqref{new3pt} 
 by diagrammatic calculation and effective potential analysis. 
This is the first example of WL scalar mass in flux compactification. 
\subsection{Set up}

The Lagrangian we consider is given by \eqref{gaugelag}, \eqref{scalarlag} and \eqref{new3pt},
	\begin{align}
	\mathcal{L}=-\frac{1}{4}F_{MN}F^{MN}-D_M\overline{\Phi}D^M\Phi+\kappa(\bar{\phi}\overline{\Phi}\Phi+\phi\overline{\Phi}\Phi),
	\end{align}
where $\kappa$ is a dimensionless coupling constant.
Using the expansion of $\phi=\braket{\phi}+\varphi$,
the Lagrangian is deformed as
	\begin{align}
	\mathcal{L}&\supset-\frac{1}{4} F^{\mu \nu} F_{\mu \nu}
	-D_{\mu} \overline{\Phi} D^{\mu} \Phi-m^2_{scalar}\bar{\Phi}\Phi \nonumber \\
	&\quad- i g \sqrt{2\alpha} \bar{\varphi} \bar{\Phi} a^{\dagger} \Phi+ i g \sqrt{2\alpha} \varphi \bar{\Phi} a \Phi
	-2 g^{2} \bar{\varphi} \varphi \overline{\Phi} \Phi \nonumber \\
	&\quad+\kappa(\bar{\varphi}\overline{\Phi}\Phi+\varphi\overline{\Phi}\Phi)+\kappa(\braket{\bar{\phi}}
	\overline{\Phi}\Phi+\braket{\phi}\overline{\Phi}\Phi),
	\end{align}
where we note that the unnecessary terms are omitted.
To derive a four-dimensional effective Lagrangian by KK reduction, 
we need to expand $\Phi$ in terms of mode functions $\xi_{n,j}$
	\begin{align}
	\Phi=\sum_{n,j}\Phi_{n,j}\xi_{n,j}. \label{phiKKexpand}
	\end{align}
Integrating over $T^2$, the four-dimensional effective Lagrangian is obtained by
	\begin{align}
	\mathcal{L}_{4D}&=-\frac{1}{4} F^{\mu \nu} F_{\mu \nu}-\partial^{\mu} \bar{\varphi} \partial_{\mu} \varphi \nonumber \\
	&\quad+\sum_{n, j}\left(-D_{\mu} \overline{\Phi}_{n, j} D^{\mu} \Phi_{n, j}-\alpha\left(n+\frac{1}{2}\right)
	\overline{\Phi}_{n, j} \Phi_{n, j}\right. \nonumber \\
	&\quad- i g \sqrt{2\alpha(n+1)} \overline{\varphi} \overline{\Phi}_{n+1, j} \Phi_{n, j}+ i g \sqrt{2\alpha(n+1)} 
	\varphi \overline{\Phi}_{n, j} \Phi_{n+1, j}-2g^2\bar{\varphi}\varphi\overline{\Phi}_{n,j}\Phi_{n,j} \nonumber \\
	&\quad+\kappa\bar{\varphi}\overline{\Phi}_{n,j}\Phi_{n,j}+\kappa\varphi\overline{\Phi}_{n,j}\Phi_{n,j}+\kappa\braket{\phi}_I
	\overline{\Phi}_{n,j}\Phi_{n,j}+\kappa\braket{\bar{\phi}}_I\overline{\Phi}_{n,j}\Phi_{n,j}\Big), \label{4Dlag}
	\end{align}
where $\braket{\phi}_I$ and $\braket{\bar{\phi}}_I$ are expressed by
	\begin{align}
	\braket{\phi}_I=\int_{T^2}dx^2\braket{\phi}\bar{\xi}_{n,j}\xi_{n',j'},~~~\braket{\bar{\phi}}_I
	=\int_{T^2}dx^2\braket{\bar{\phi}}\bar{\xi}_{n,j}\xi_{n',j'}.
	\end{align}
When $\braket{\phi}=f\bar{z}/\sqrt{2}$, $\braket{\phi}_I$ and $\braket{\bar{\phi}}_I$ lead to zero 
 because of odd function with respect to integral variables $z$ or $\bar{z}$.

\subsection{Diagrammatic computation}
Before computing a finite quantum correction, 
 we first review that the quantum correction to WL scalar mass is cancelled at one-loop in the case of $\kappa=0$ (see \cite{B1}).
Computation of Feynman diagram is expressed as
	\begin{align}
	\mathcal{I}_{4pt}&=-i2g^2\sum_{n,j}\int\frac{d^4 k}{(2\pi)^4}\frac{1}{k^2+\alpha\left(n+\frac{1}{2}\right)}, \\
	\mathcal{I}_{3pt}&=+i2g^2\sum_{n,j}\int\frac{d^4 k}{(2\pi)^4}\frac{\alpha(n+1)}{\left(k^2+\alpha\left(n+\frac{1}{2}\right)\right)
	\left(k^2+\alpha\left(n+\frac{3}{2}\right)\right)}.
	\end{align}
Note that $\mathcal{I}_{3pt}$ is the example of section \ref{differentKK}.
The sum of $\mathcal{I}_{4pt}$ and $\mathcal{I}_{3pt}$ is obtained by
	\begin{align}
	\mathcal{I}_{4pt}+\mathcal{I}_{3pt}=-2ig^2|N|\sum_{n}\int\frac{d^4 k}{(2\pi)^4}
	\left(\frac{n+1}{k^2+\alpha\left(n+\frac{3}{2}\right)}-\frac{n}{k^2+\alpha\left(n+\frac{1}{2}\right)}\right). \label{cancelcal}
	\end{align}
By the shift $n\rightarrow n+1$ in the second term, \eqref{cancelcal} becomes zero.

For $\kappa\ne0$, we get a new quantum correction to WL scalar mass from the right diagram in figure \ref{fig:one}.
Computing the diagram results in
	\begin{align}
	\mathcal{I}
	&=+i\kappa^2\sum_{n, j} \int \frac{d^{4} k}{(2 \pi)^{4}} \frac{1}{\left(k^{2}+\alpha\left(n+\frac{1}{2}\right)\right)^2}
	=\frac{i\kappa^2|N|}{\alpha^2}\left(\frac{4\pi}{\alpha}\right)^{\epsilon-2}J(1/2;0,2).
	\label{Inew}
	\end{align}
Applying $J(1/2;0,2)$ to \eqref{Jxa2}, we obtain $J(1/2; 0,2)=\Gamma(\epsilon)\zeta[\epsilon,1/2]$ and then
	\begin{align}
	\mathcal{I}&=\frac{i\kappa^2|N|}{\alpha^2}\left(\frac{4\pi}{\alpha}\right)^{\epsilon-2}\Gamma(\epsilon)\zeta[\epsilon,1/2] \nonumber \\
	&=-i\frac{\kappa^2|N|\ln2}{32\pi^2}\left(\frac{4\pi}{\alpha}\right)^\epsilon+\mathcal{O}(\epsilon),
	\label{explicitQC}
	\end{align}
where
	\begin{align}
	\Gamma(\epsilon)&=\frac{1}{\epsilon}-\gamma+\mathcal{O}(\epsilon), \\
	\zeta[\epsilon,1/2]&=0-\epsilon\frac{\ln2}{2}
	\end{align}
are used in the second line of \eqref{explicitQC}.
This correction is finite in $\epsilon\rightarrow 0$ limit.
Thus, the quantum correction to WL scalar mass at one-loop is given by
	\begin{align}
	\delta m^2=i\mathcal{I}=\frac{|N|\ln2}{32\pi^2}\frac{\kappa^2}{L^2}. 
	\label{FDview}
	\end{align}
Note that we introduced a factor of torus area $L^2$, which comes from the normalization factors for KK mode function.
Obviously, $\delta m^2=0$ is reproduced for $\kappa=0$ in six-dimensional scalar QED (see \cite{B1}).
One of the interesting phenomenological applications is that the quantum correction $\delta m^2$ to WL scalar mass 
 can be interpreted as Higgs mass. 
This idea is based on gauge-Higgs unification, namely a zero-mode of $\varphi$ is regarded as Higgs field.
Even if the compactification scale is Planck scale $1/L\sim\mathcal{O}(M_{Planck})$, 
 Higgs mass could be realized by the interaction term \eqref{new3pt} generated 
 by some dynamics at $\mathcal{O}$(1) TeV scale.
This is analogous to the mass of pion as a pseudo NG boson for chiral symmetry. 
The reason why the pion mass is not Planck scale is that chiral symmetry is dynamically broken at extremely lower energy scale 
 comparing to the Planck scale, namely, QCD scale. 
Note that the similar discussion cannot be applied to the ordinary gauge-Higgs unification 
since $\kappa$ is replaced by an $SU(2)_L$ gauge coupling in this scenario.


\subsection{Effective potential analysis}
Next, we consider the quantum correction to WL scalar mass in terms of effective potential.
In our setup \eqref{4Dlag}, we read the KK mass spectrum of $\Phi$ to be 
 $\alpha(n+1/2)-\kappa\braket{\phi}_I-\kappa\braket{\bar{\phi}}_I$.
Thus, the four-dimensional effective potential is given by
	\begin{align}
	V=\sum_{n=0}^\infty\int \frac{d^{4} k}{(2 \pi)^{4}}\ln\left(k^2+\alpha\left(n+\frac{1}{2}\right)
	-\kappa\braket{\phi}_I-\kappa\braket{\bar{\phi}}_I\right),
	\end{align}
where we take into account a degree of freedom of complex scalar field $\Phi$.
To obtain quantum correction to WL scalar mass from four-dimensional effective potential, 
 we differentiate effective potential with respect to $\braket{\phi}_I$ and $\braket{\bar{\phi}}_I$.
Thus, $\delta m^2$ is obtained as
	\begin{align}
	\delta m^2&=\left.\frac{\partial^2V}{\partial\braket{\phi}_I
	\partial\braket{\bar{\phi}}_I}\right|_{\braket{\phi}_{I}=0} \nonumber \\
	&=-\kappa^2\sum_{n=0}^\infty\int \frac{d^{4} k}{(2 \pi)^{4}}
	\frac{1}{\left(k^2+\alpha\left(n+\frac{1}{2}\right)\right)^2}=i\mathcal{I}. 
	\label{EPview}
	\end{align}
This result \eqref{EPview} agrees with \eqref{Inew} or \eqref{FDview}.

\subsection{Explicit breaking of translational invariance in extra space}
Before confirming explicit breaking of translational invariance in extra space, 
 we review NG boson of translational invariance in extra space in the case of $\kappa=0$.
If $\kappa=0$, the quantum correction to WL scalar mass becomes zero.
This physical reason can be understood from the fact 
 that the zero-mode of $\varphi$ is a NG boson of translational invariance in extra space.
The breaking of translational invariance by the background $\braket{\phi}$ can be replaced for a shift of $\varphi$,
	\begin{align}
	\delta_T \varphi=(\epsilon\partial+\bar{\epsilon}\bar{\partial})\varphi+\frac{\bar{\epsilon}}{\sqrt{2}}f,
	\end{align}
where $\epsilon\equiv(\epsilon_5+i\epsilon_6)/2$ and 
 $\epsilon_5,~\epsilon_6$ means constant parameters of translations in extra spaces.
Focusing on the zero-mode of $\varphi$ and noticing $\partial\varphi=\bar{\partial}\varphi=0$,
we obtain
	\begin{align}
	\delta_T\varphi=\frac{f}{\sqrt{2}}\bar{\epsilon}.
	\end{align}
This shows that the zero-mode of $\varphi$ is identified with a NG boson under the translation in extra spaces.

If $\kappa\ne0$, $\varphi\overline{\Phi}_{n,j}\Phi_{n,j}$ (or $\bar{\varphi}\overline{\Phi}_{n,j}\Phi_{n,j}$) in \eqref{4Dlag} 
 is expected to break the translational invariance.
To confirm it, we consider the following local six-dimensional transformation \cite{B2}
	\begin{align}
	\varphi'=\varphi-\frac{1}{\sqrt{2}}\partial\Lambda,~~~\Phi'=e^{g\Lambda}\Phi,~~~\overline{\Phi}'=e^{-g\Lambda}\overline{\Phi},
	\label{gaugetrf}
	\end{align}
where $\Lambda=f(\epsilon\bar{z}-\bar{\epsilon}z)$.
Infinitesimal transformation of $\epsilon,\bar{\epsilon}$ is expressed as
	\begin{align}
	\delta_\Lambda\varphi=-\frac{1}{\sqrt{2}}\partial\Lambda,~~~\delta_\Lambda\Phi
	=g\Lambda\Phi,~~~\delta_\Lambda\overline{\Phi}=-g\Lambda\overline{\Phi}.
	\end{align}
Transformations of $\varphi$ and $\Phi$ are the combination of translation $\delta_T$ 
 and infinitesimal transformation $\delta_\Lambda$,
	\begin{align}
	\delta\varphi&=(\delta_T+\delta_\Lambda)\varphi=\sqrt{2}f\bar{\epsilon}, \label{WLtrans}\\
	\delta\Phi&=(\delta_T+\delta_\Lambda)\Phi=-i\sqrt{\alpha}(\epsilon a^\dag+\bar{\epsilon}a)\Phi.
	\end{align}
Using \eqref{creatannihilate} and \eqref{phiKKexpand}, we obtain
	\begin{align}
	\delta\Phi&=-i\sqrt{\alpha}\sum_{n,j}\Phi_{n,j}(\epsilon a^\dag+\bar{\epsilon}a)\xi_{n,j}=\sum_{n,j}\delta\Phi_{n,j}\xi_{n,j}, \\
	\delta\Phi_{n,j}&=-i\sqrt{\alpha}(\epsilon\sqrt{n+1}\Phi_{n+1,j}+\bar{\epsilon}\sqrt{n}\Phi_{n-1,j}). \label{phinj}
	\end{align}
For $\delta\overline{\Phi}_{n,j}$, it is given by complex conjugate of \eqref{phinj},
	\begin{align}
	\delta\overline{\Phi}_{n,j}&=+i\sqrt{\alpha}(\bar{\epsilon}\sqrt{n+1}\overline{\Phi}_{n+1,j}
	+\epsilon\sqrt{n}\overline{\Phi}_{n-1,j}). \label{barphinj}
	\end{align}

Let us confirm the explicit breaking of translational invariance of the interaction term $\varphi\overline{\Phi}_{n,j}\Phi_{n,j}$.
First, a transformation of $\overline{\Phi}_{n,j}\Phi_{n,j}$ is 
	\begin{align}
	\delta\left(\sum_{n,j}\overline{\Phi}_{n,j}\Phi_{n,j}\right)&=i\sqrt{\alpha}\sum_{n,j}\Big(\bar{\epsilon}\sqrt{n+1}
	\overline{\Phi}_{n+1,j}\Phi_{n,j}+\epsilon\sqrt{n}\overline{\Phi}_{n-1,j}\Phi_{n,j} \nonumber \\
	&\quad\quad\quad\quad\quad-\epsilon\sqrt{n+1}\overline{\Phi}_{n,j}\Phi_{n+1,j}-\bar{\epsilon}\sqrt{n}
	\overline{\Phi}_{n,j}\Phi_{n-1,j}\Big) \nonumber \\
	&=0.
	\end{align}
Thus, the mass term of $\Phi_{n,j}$ is invariant.
For $\varphi\overline{\Phi}_{n,j}\Phi_{n,j}$, a transformation is
	\begin{align}
	\delta\left(\sum_{n,j}\varphi\overline{\Phi}_{n,j}\Phi_{n,j}\right)&=(\delta\varphi)
	\sum_{n,j}\overline{\Phi}_{n,j}\Phi_{n,j}+\varphi\delta\left(\sum_{n,j}\overline{\Phi}_{n,j}\Phi_{n,j}\right) \nonumber \\
	&=\sqrt{2}f\bar{\epsilon}\sum_{n,j}\overline{\Phi}_{n,j}\Phi_{n,j}\ne0.
	\end{align}
This result means the explicit breaking of translational invariance in extra spaces.
For $\kappa\ne0$, the zero-mode of $\varphi$ is identified with a pseudo NG boson of translational invariance in extra spaces.

One might claim that the interaction terms \eqref{new3pt} are not gauge invariant 
 since $\varphi$ or $\overline{\varphi}$ transforms under the gauge symmetry as (\ref{gaugetrf}). 
In ordet to overcome such a claim, 
 $\varphi$ or $\overline{\varphi}$ should be expressed by a gauge invariant non-local Wilson line operator 
  and the interaciton terms \eqref{new3pt} should be regarded 
  as one of the terms of expanding the Wilson line operators in small $\varphi$ or $\overline{\varphi}$. 
Noting that the Wilson line operators\footnote{In non-Abelian case, 
the path ordered operation must be taken into account, $U_{5,6} = P \exp [ig \oint A_{5,6} dx^{5,6}]$.} 
\begin{align}
U_5 = \exp \left[ ig \oint A_5 dx^5 \right], \quad U_6 = \exp \left[ ig \oint A_6 dx^6 \right]
\end{align}
can be written in terms of $\varphi, \overline{\varphi}$ and $z, \bar{z}$ as
\begin{align}
&U_5 = \exp \left[ 
\frac{g}{\sqrt{2}} \oint 
(\varphi dz + \varphi d\bar{z} - \overline{\varphi} dz - \overline{\varphi} d\bar{z}) 
\right], \\
&U_6 = \exp \left[ \frac{g}{\sqrt{2}} \oint 
(\varphi dz - \varphi d\bar{z} + \overline{\varphi} dz - \overline{\varphi} d\bar{z}) 
\right],
\end{align}
we find that the cubic terms introduced in this paper can be expressed 
by the non-local Wilson line operators
\begin{align}
i(U_5-U_5^\dag) \overline{\Phi} \Phi - i(U_6-U_6^\dag) \overline{\Phi} \Phi 
&\supset 2\sqrt{2} i g \oint \varphi d\bar{z} \overline{\Phi} \Phi
-2\sqrt{2} i g \oint \overline{\varphi} dz \overline{\Phi} \Phi 
\nonumber \\
&=
2\sqrt{2} i g_4  \varphi \overline{\Phi} \Phi
-2\sqrt{2} i g_4 \overline{\varphi} \overline{\Phi} \Phi
\label{WLcubic}
\end{align}
 where $g_4$ is a four dimensional gauge coupling constant. 
Note that the $\overline{\Phi} \Phi$ term cannot be included in (\ref{WLcubic}). 
If this term is allowed, the WL scalar mass would be divergent. 

We comment on how the finite WL scalar mass can be expressed in terms of the Wilson line operators. 
If the WL scalar mass is generated in the broken phase, 
 where the VEV of the WL scalar field is non-zero, 
 it is straightforward to express the WL scalar mass by the Wilson line operators 
 as in the gauge-Higgs unification. 
As for the WL scalar field mass in the present paper, 
 the mass is generated in the unbroken phase and is independent of the VEV of the WL  scalar field. 
Therefore, we cannot express the WL scalar field mass by the Wilson line operators explicitly.  

Under the constant shift of $A_5 \to A_5 - f \epsilon_6/2, A_6 \to A_6 + f \epsilon_5/2$, 
 the operators 
\begin{align}
U_5 - U_5^\dag = 2i \sin \left[ g \oint A_5 dx^5 \right], \quad 
U_6 - U_6^\dag = 2i \sin \left[ g \oint A_6 dx^6 \right]
\end{align}
are not obviously invariant, 
 which means that the interaction terms (\ref{WLcubic}) explicitly break the shift symmetry.   
Clarifying the origin of the interaction terms (\ref{WLcubic}) is not easy and beyond the scope of this paper. 
We expect that the origin of the interaction terms would be connected to the quantum gravity effects, 
nontrivial backgrounds such as a vortex, or some non-perturbative dynamics. 
We leave this issue for our future work. 


\section{Conclusion and Discussion}
We have studied the possibility to realize nonvanishing WL scalar mass in flux compactification.
Using KK mass of various fields in the bulk, 
 we have generalized loop integrals in the quantum correction to the WL scalar mass and
have systematically analyzed their structure of divergence. 
The conditions for the loop integral and the mode sum to be finite were derived. 
Then, we have classified four-point and three-point interaction terms 
 providing finite quantum corrections to WL scalar mass at one-loop.


Of these interaction terms, we focused on $\bar{\varphi}\overline{\Phi}\Phi+\varphi\overline{\Phi}\Phi$ type interaction 
 since these interaction terms have no derivatives and are the simplest of all interaction terms. 
Using these interactions in a six dimensional scalar QED, 
 we have explicitly illustrated the nonvanishing finite quantum correction to WL scalar mass in two ways: 
  diagrammatic computation or effective potential analysis, 
  which is the first example of finite WL scalar mass in flux compactification.
This result is easy to understand, which analogous to the case of pion. 
WL scalar was originally a NG boson of translational symmetry in extra space and therefore massless. 
Introducing an interaction term $\varphi\overline{\Phi}\Phi$ of explicit breaking of translational symmetry 
 made WL scalar into a pseudo NG boson and WL scalar obtain a finite mass.  


Phenomenologically, our result can be applied to Higgs physics. 
If we regard \eqref{FDview} as Higgs mass, $\kappa/L \approx \mathcal{O}(1)$ TeV. 
If the interactions explicitly breaking translational invariance are generated around TeV scale by some dynamics, 
 Higgs mass will be expected to be obtained even if the compactification scale is an order of the Planck scale.  
Extension to a realistic model of electroweak symmetry breaking is very interesting to consider in future study. 


\section*{Acknowledgments}
This work is supported in part by JSPS KAKENHI Grant Number JP17K05420 (N.M.).

\appendix
\section{The property of Hurwitz zeta function}\label{HZF}

Hurwitz zeta function is given by
	\begin{align}
	\zeta[s,a]=\sum_{n=0}^\infty\frac{1}{(n+a)^s}.
	\end{align}
It is known that Hurwitz zeta function is related to Riemann zeta function by the following identical equation
	\begin{align}
	\zeta[s,1]&=\zeta(s), \\
	\zeta[s,1/2]&=(2^s-1)\zeta(s).
	\end{align}
Since Riemann zeta function satisfies $\zeta(-2n)=0$ ($n$ is a positive integer), Hurwitz zeta function also satisfies
	\begin{align}
	\zeta[-2n,1]=0,~~~\zeta[-2n,1/2]=0. \label{finitekey}
	\end{align}
In particular, $\zeta[s,1/2]$ is satisfied by $\zeta[0,1/2]=0$.


\end{document}